\documentclass{egpubl_pg_short}
\usepackage{pg2018s}
\WsPaper
\electronicVersion

\ifpdf \usepackage[pdftex]{graphicx} \pdfcompresslevel=9
\else \usepackage[dvips]{graphicx} \fi

\PrintedOrElectronic
\usepackage{t1enc,dfadobe}
\usepackage{egweblnk}
\usepackage{cite}

\usepackage{balance}  
\usepackage{graphics} 
\usepackage{times}    
\usepackage{url}      

\usepackage{multicol}
\usepackage{cuted}
\usepackage[font=small,labelfont=bf,tableposition=top]{caption}
\newcommand{\tool}{Tabby}
\title{\tool{}: Explorable Design for 3D Printing Textures}

\author[R. Suzuki, K. Yatani, M. D. Gross \& T. Yeh]
{
\parbox{\textwidth}{\centering R. Suzuki$^{1}$, K. Yatani$^{2}$, M. D. Gross$^{1}$, and T. Yeh$^{2}$}
\\
{\parbox{\textwidth}{\centering $^1$University of Colorado Boulder\\ $^2$The University of Tokyo}}
\vspace{-1.0cm}
}

\begin{document}

\teaser{
 \includegraphics[width=\linewidth]{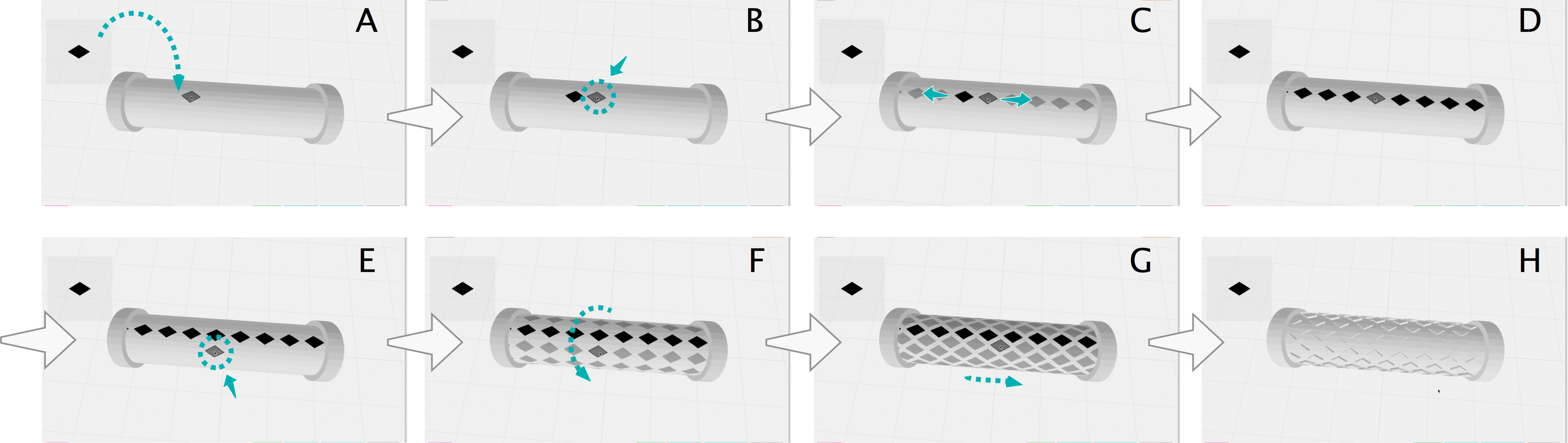}
 \centering
\captionof{figure}{An overview of the workflow for creating gripping surface texture with Tabby. (A) Users sketch a texture element for grips and drag-and-drop it onto the 3D model. The system infers a possible target surface based on the user interaction. (B) Users copy and paste the texture element a couple of times. (C) When the system detects repetitive operations, Tabby suggests texture patterns. (D) Users can accept the suggested pattern. (E -- G) Another example of texture creation with Tabby. (H) Users extrude the texture, creating a 3D-printable pattern.}~\label{fig:workflow}
}

\maketitle

\begin{abstract}
This paper presents \tool{}, an interactive and explorable design tool for 3D printing textures.
\tool{} allows texture design with direct manipulation in the following workflow: 1) select a target surface, 2) sketch and manipulate a texture with 2D drawings, and then 3) generate 3D printing textures onto an arbitrary curved surface.
To enable efficient texture creation, \tool{} leverages an auto-completion approach which automates the tedious, repetitive process of applying texture, while allowing flexible customization.
Our user evaluation study with seven participants confirms that \tool{} can effectively support the design exploration of different patterns for both novice and experienced users.



\end{abstract}

\section{Introduction}
Texture is an essential property of physical objects that affects aesthetics, usability, and functionality. 
For example, patterns in a lampshade can enhance the aesthetics of the light design, brick patterns in architectural models can improve visual details, and a rough texture can affect an object's usability by providing a gripping surface.
Designing and applying textures, however, remains difficult and time-consuming.
It requires proficient 3D modeling skills and manual repetitive operations. 
Existing CAD tools usually support texture design through parametric modeling which allows a designer to explore different patterns through changing parameters or to synthesize textures from pre-defined examples.
Such parametric design is powerful, but it suffers from two key limitations.
First, the design process involves indirect manipulation of repetitive parameter tuning, which creates a large gulf of execution.
Second, there is a steep learning curve to master commands.
Prior work reveals that typical CAD users do not utilize most of the available commands in such tools due to high learning cost~\cite{matejka2009communitycommands}.
Thus, our formative study reveals many designers, even experienced users, often simply repeat copy-and-paste to create desired patterns, and it severely limits flexibility of design exploration.
This observation leads us to explore an alternative interaction model where users can naturally adopt high-level commands and explore alternative designs through direct manipulation.

To investigate this new approach, we developed \tool{}, an interactive design tool for creating and exploring 3D printing textures.
With \tool{}, a designer would only need to demonstrate the first few units of a texture pattern.
Then, the system can automatically infer a complete pattern the designer may have in mind.
The designer can simply accept it or adjust to achieve the desired texture (Figure~\ref{fig:workflow}).
Informed by our formative study, we introduce three techniques for efficient texture creation:
1) \emph{auto-completion}: given the user's copy and paste operation, our system automates the tedious, repetitive process of applying texture, 
2) \emph{2D element manipulation}: in contrast to existing 3D modeling tools, our system supports 2D drawing operations to allow the user to define and arrange a texture element which is later automatically converted to a 3D printable texture
, and 
3) \emph{semantic region selection}: our system infers which surface region to fill the repeated pattern.

To evaluate the efficiency and flexibility of texture design with \tool{}, we conducted a controlled experiment with seven designers.
Our study shows that \tool{} speeds up texture creation by 80\% over conventional tools.
This performance gain becomes even larger with more complex target surfaces. 
Our qualitative result confirms that designing and applying textures with our system is simpler and more effective.

In summary, this paper contributes: 
\begin{enumerate}
\item \tool{}, an interactive system that instantiates the auto-completion method in the context of 3D texture design;
\item A set of workflow and techniques that support 2D operations for designing 3D-printable texture patterns;
\item A controlled experiment with seven designers that shows users can efficiently and flexibly design textures compared to conventional tools.
\end{enumerate}

\begin{figure}[th!]
\centering
\includegraphics[width=0.9\columnwidth]{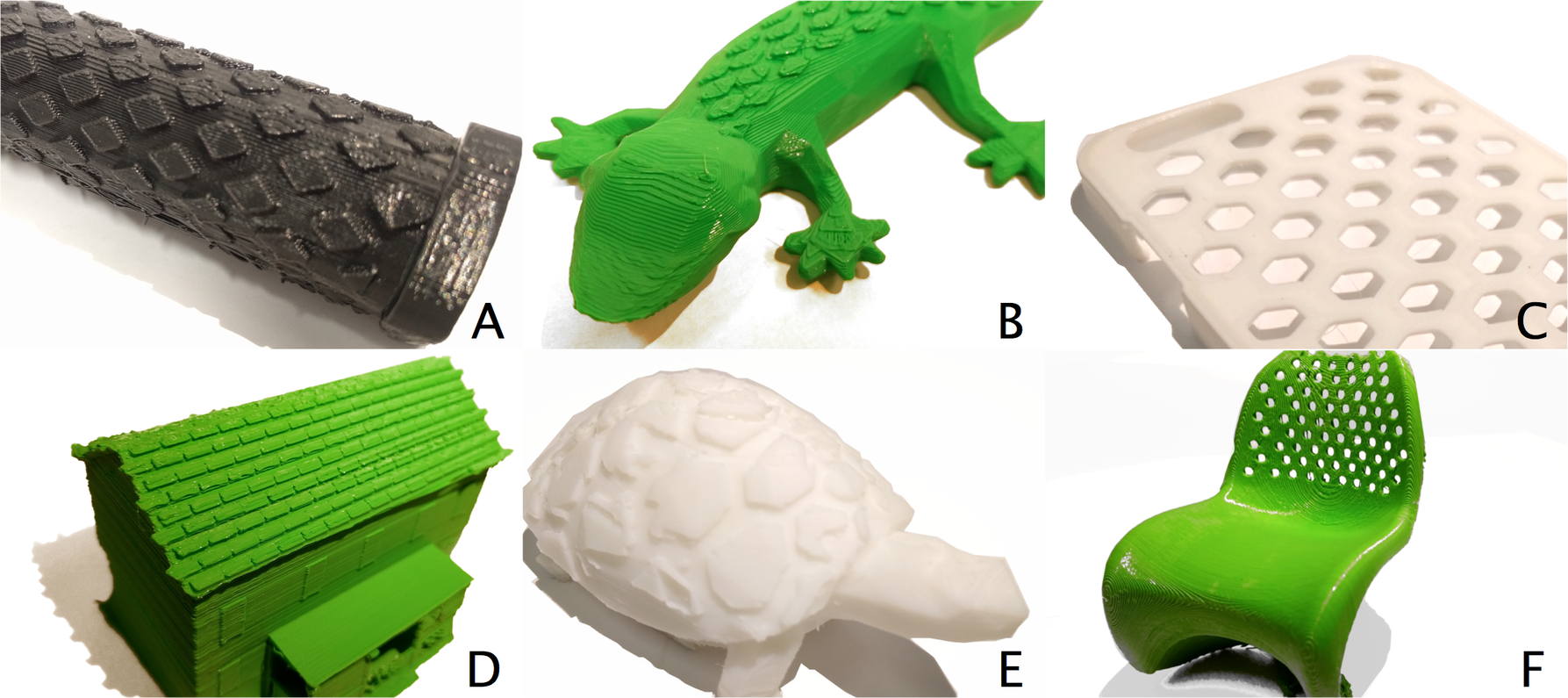}
\caption{Examples of 3D printed texture with Tabby. 
}~\label{fig:cover}
\end{figure}

\vspace{-0.3cm}
\section{Related Work}

Prior research has developed interactive systems for various mesh editing tasks.
Systems like MeshMixer \cite{schmidt2010meshmixer} and GeoBrush \cite{takayama2011geobrush} provide an interactive texture cloning tool, and users can copy a geometric feature from example models by a simple brushing operation.
However, these systems are not designed to support the creation of repeated patterns like tactile textures.
Exemplars of 3D texture are typically difficult to obtain, inducing another limitation in existing texture cloning tools.
Also, a user must manually choose a region of both the source texture and the target surface.
Such a task can be time-consuming if the target region is large.

Other approaches have demonstrated texture creation through an automatic geometric synthesis method~\cite{dumas2015example, zhou2006mesh}.
However, these systems mainly aim to fully automate texture synthesis, and do not support interactive design explorations.
One of our main objectives is to support interactive tactile texture creation for 3D models.
Similar to our work, prior systems have introduced an interactive tool for texture pattern synthesis~\cite{landreneau2010scales, torres2015hapticprint}.
While users can adjust the position and orientation of the pattern, the design exploration process in these tools is mostly limited in parameter tuning.
In contrast, Tabby enables users to naturally and intuitively explore different texture patterns through direct manipulation.

Creating repetitive patterns is a tedious manual process.
To alleviate user workload, prior work has developed auto-completion techniques in 2D drawing applications.
Kazi et al.~\cite{kazi2012vignette} demonstrated Vignette, an interactive drawing application that can facilitate user-defined 2D textures.
The user draws a part of a texture and gestures to automatically fill a 2D region with the texture.
Later research adapted this concept to enable data-driven decorative patterns~\cite{lu2014decobrush}, hand-drawn patterns~\cite{xing2014autocomplete,xing2015autocomplete}, and 3D sculpting~\cite{peng2018autocomplete}.
These systems inspired us to investigate how we can integrate these auto-complete techniques into 3D modeling to support interactive texture design.

\vspace{-0.3cm}
\section{Design Goals}

To discover the needs and challenges, we conducted two formative studies where we observed the texture design process of 25 novice users and 3 professional CAD users.
These observations led us to the following high-level design goals.

\begin{figure}[ht!]
\centering
\includegraphics[width=\columnwidth]{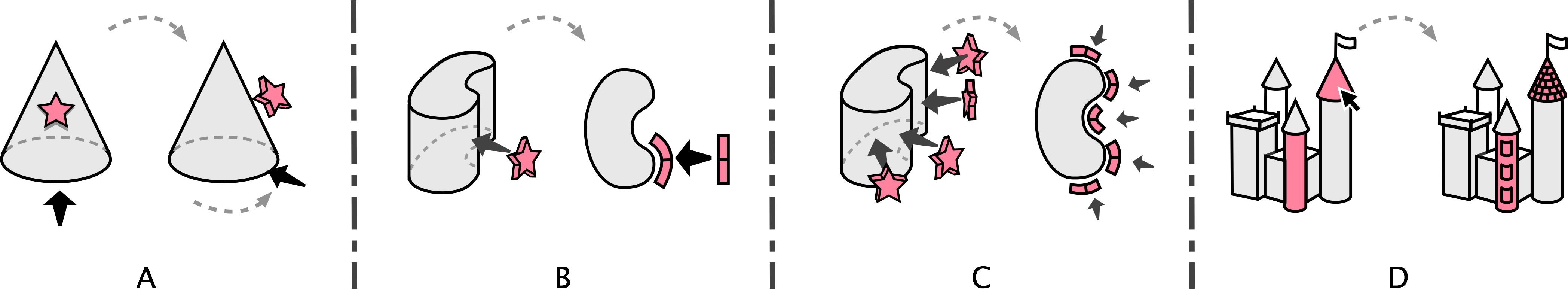}
\caption{Four difficulties in texture design revealed by the formative study.
}~\label{fig:design-goals}
\end{figure}

\begin{enumerate}
\item \textbf{Exploiting 2D Operations:} We found that novice users struggle to align, locate, and manipulate objects in 3D space.
They needed to continuously change the camera angle to ensure that all texture elements were attached correctly to the target surface (Figure~\ref{fig:design-goals}A).
This task becomes even harder on curved surfaces because the user must deform a texture to fit the target curvature (Figure~\ref{fig:design-goals}B).
In contrast, 3D modeling becomes manageable when it is presented as a series of 2D operations~\cite{hudson2016understanding}.
Thus, we decided to exploit 2D operations in our system.
\item \textbf{Liberating from Repetitive Operations:}
Textures often consist of repeated patterns.
We observed that users typically perform copy-and-paste operations to create such repetitive patterns.
Thus, a repetition of such operations can signal an attempt to create textures, particularly for novice users.
While it is a natural workflow, this simple copy-and-paste approach can be tedious since it requires many repetitive operations (Figure~\ref{fig:design-goals}C).
To simplify this manual operation, we should exploit this behavior to infer the user's intentions and, if possible, automatically complete the intended texture.
\item \textbf{Allowing Intuitive Exploration:}
Texture design process is not one time.
Users usually need to explore different patterns by changing the design properties (e.g., shape, size, rotation, orientation, alignment and region to fill).
While users often change these properties through parameter tuning, we found that it becomes difficult when designers have to deal with multiple parameters and constraints simultaneously.
Thus, our tool should support designers to easily map their actions to intended results through direct manipulation (Figure~\ref{fig:design-goals}D).
\end{enumerate}

\vspace{-0.3cm}
\section{Tabby: An Interactive and Explorable Texture Design Tool}

Our system, Tabby, supports the design process of rich, user-defined tactile texture creation.
The typical workflow in Tabby is as follows:

\begin{description}
\item[\emph{Step 1:}] Import a 3D object into Tabby's working space.
\item[\emph{Step 2:}] Sketch or import an SVG image for a desired texture element, and place it on the object (Figure~\ref{fig:workflow}A).
\item[\emph{Step 3:}] The system infers the target region based on the placement of the texture element. Users can accept the suggested pattern (Figure~\ref{fig:workflow}D) or adjust the region selection if needed.
\item[\emph{Step 4:}] After users copy and paste the texture elements a couple of times, the system suggests auto-completion of the pattern  (Figure~\ref{fig:workflow}B, C, E and F).
\item[\emph{Step 5:}] Users adjust the properties of the pattern (Figure~\ref{fig:workflow}G).
\item[\emph{Step 6:}] Once users confirm, the system extrudes the 2D texture, converting it into a 3D geometry while maintaining water-tightness (Figure~\ref{fig:workflow}H).
\item[\emph{Step 7:}] Users can download the modified model for 3D printing. 
\end{description}


\vspace{-0.3cm}
\subsection{Semantic Region Selection}
\begin{figure}[!h]
\centering
\includegraphics[width=0.8\columnwidth]{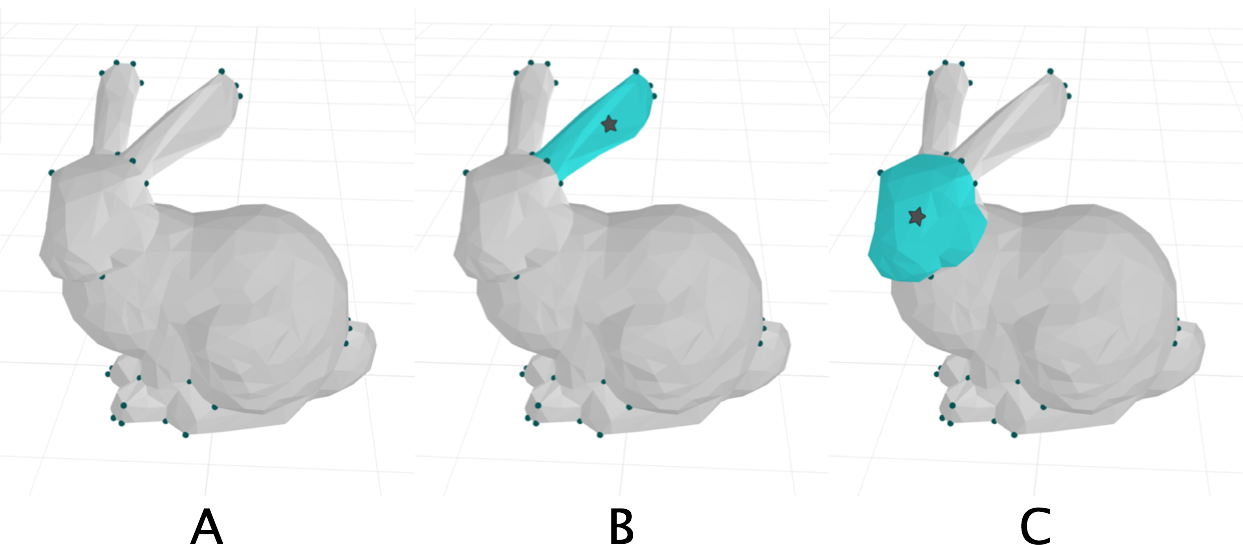}
\caption{Semantic region selection based on the cursor position.
}~\label{fig:region-selection}
\end{figure}

Users start with defining an element for texture patterns.
In Tabby, they draw the element in a 2D sketching canvas or import the element as an SVG file.
After deciding the element, users drag it into the main working space.
The system displays a shadow of the element as visual feedback.
As users move the element, the system automatically infers the surface region where they intend to create textures, and highlights it in light blue (Figure~\ref{fig:region-selection}).

To enable the interactive region selection, we leverage and extend the existing mesh segmentation algorithms.
Our method is based on cross-boundary mesh decomposition~\cite{zheng2010mesh}, which computes a harmonic field with a Laplacian matrix to obtain segmentation boundaries by cutting along an isoline of the harmonic field.
To obtain boundary positions, we calculate highly distorted vertices as boundary points because the distortion occurs in high surface curvature.
The distortion of a vertex $i$ is defined as: $D (i) = \max_{0 \le r \le R} \frac{2 \pi - \sum_{j} \tau_j (r) }{2 \pi}$, where $R$ is a region radius and $\tau_j (r)$ are the angles at $i$ of face $j$ inside of region radius $r$.
After calculating the distortion of each vertex, we extract high distorted points using a terminal vertex selection algorithm~\cite{sheffer2002seamster} with weights defined by the distance between the current mouse position and the target vertices.

\vspace{-0.3cm}
\subsection{Texture Auto-completion}
\begin{figure}[th!]
\centering
\includegraphics[width=1\columnwidth]{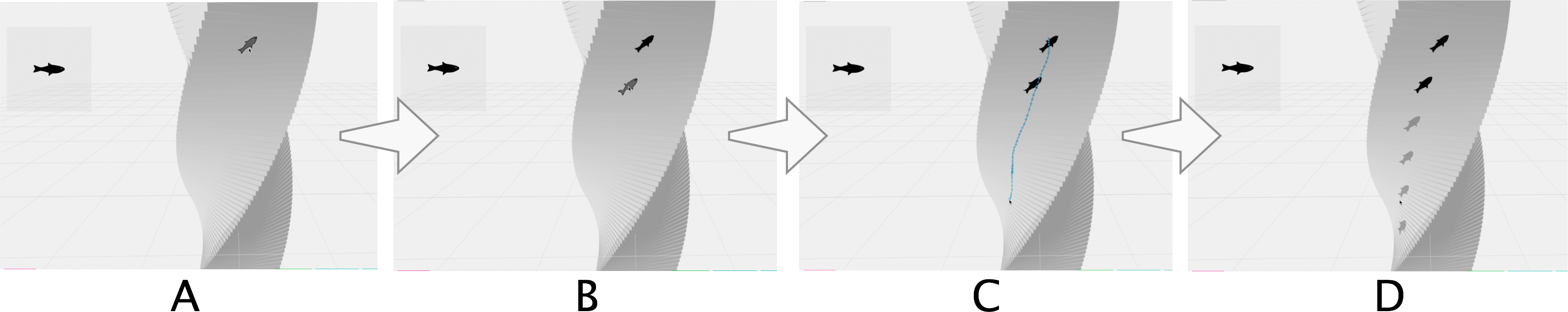}
\caption{Auto-completion of repeated patterns by drawing a line.
}~\label{fig:auto-complete}
\end{figure}

After users place the first element, they can perform copy-and-paste operations to start forming a texture by auto-completing repetitive patterns.
Tabby's auto-completion process is as follows.
First the system tries to detect the user's copy-and-paste operations.
When such operations are detected, it tracks the placements of the first two texture units and calculates the relative positions between the two.
Then, Tabby makes a suggestion for auto-completion by presenting an example where each individual element is visualized as a shadow cast on the surface (Figure~\ref{fig:workflow} and ~\ref{fig:auto-complete}).
Tabby's extrapolation can support both patterns in x and y, and curves lines with the user's additional demonstration.

\begin{figure}[h!]
\centering
\includegraphics[width=0.9\columnwidth]{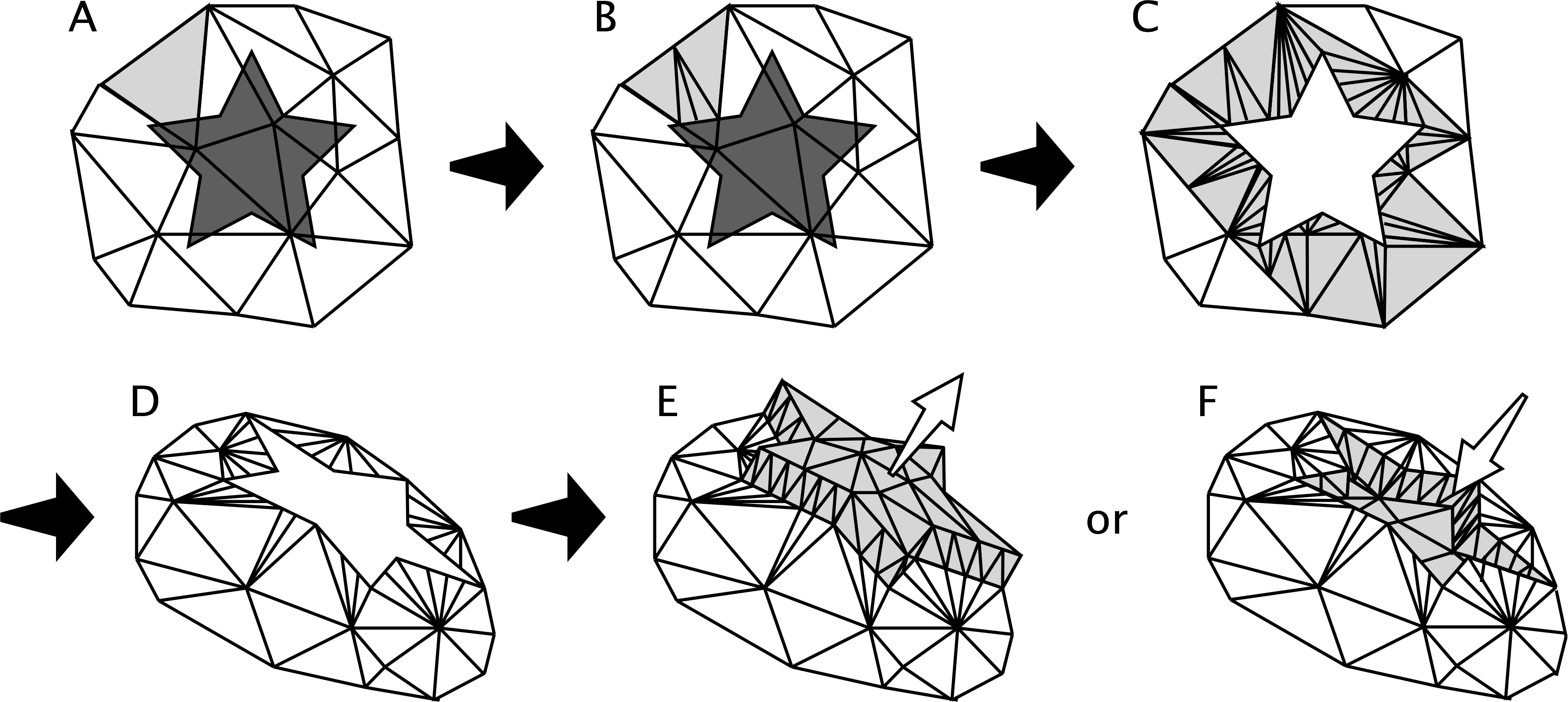}
\caption{Texture mesh generation. Triangulate the texture and replace the surfaces with the new mesh (A-D) . The user can perform extrusion or intrusion of the texture element based on the obtained vertex normals (E-F).}~\label{fig:mesh}
\end{figure}

After completing the design, the system automatically converts the 2D drawing element into a series of triangle meshes to create 3D textures.
Users can also interactively change texture types (e.g., bumps or cutting holes).
As we will show in the following section, the system also ensures that these added triangle meshes are properly fused into the target to obtain a water-tight result.
We first triangulate the texture unit, and obtain the boundary points in 2D coordinates and corresponding triangle faces  (Figure~\ref{fig:mesh}A-B).
Then, we replace the original mesh surface with triangulated new meshes (Figure~\ref{fig:mesh} C).

Once the boundary positions and surrounding surfaces are determined, the system creates a corresponding enclosure to maintain the water-tightness.
We compute the vertex normal vector for each boundary point, and get a set of outer vertices by extending with the normal vector.
We then perform the same constrained Delaunay triangulation on the outer vertices to obtain outer polygonal surfaces (Figure~\ref{fig:mesh}D).
Note that the resulting textures are smoothly deformed to fit the target curvature.
With the same interactive operation, users can also obtain an embossed texture by intruding the boundary points or a hollow texture by creating internal cavities inside the mesh with a consistent wall thickness (Figure~\ref{fig:mesh}E-F).

\begin{figure}[ht!]
\centering
\includegraphics[width=\columnwidth]{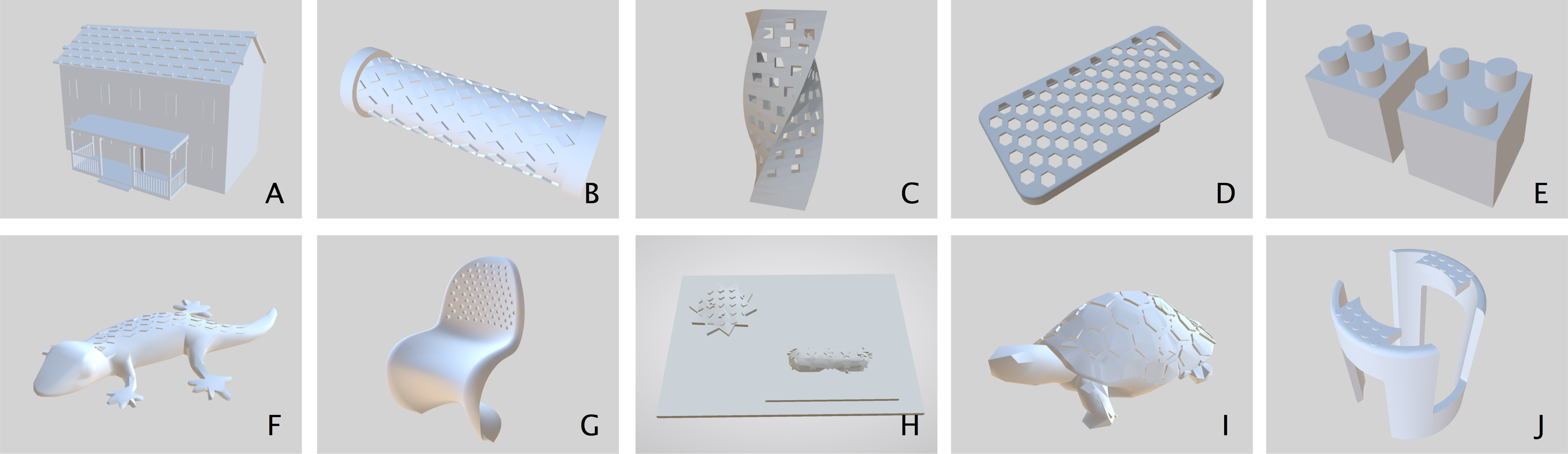}
  \caption{Examples of possible real-world use scenarios.}~\label{fig:applications}
\end{figure}

Figure~\ref{fig:applications} shows example results of texture patterns using \tool{}: A) an architectural model, B) a grip on a bike handle, C) a lampshades, D) smartphone cases, E) blocks, F) scale of geckos, G) chairs, H) tactile picture books, I) shell of turtles, and J) anti-slip finishing for cup holders.

\begin{figure}[b!]
\centering
\includegraphics[width=0.9\columnwidth]{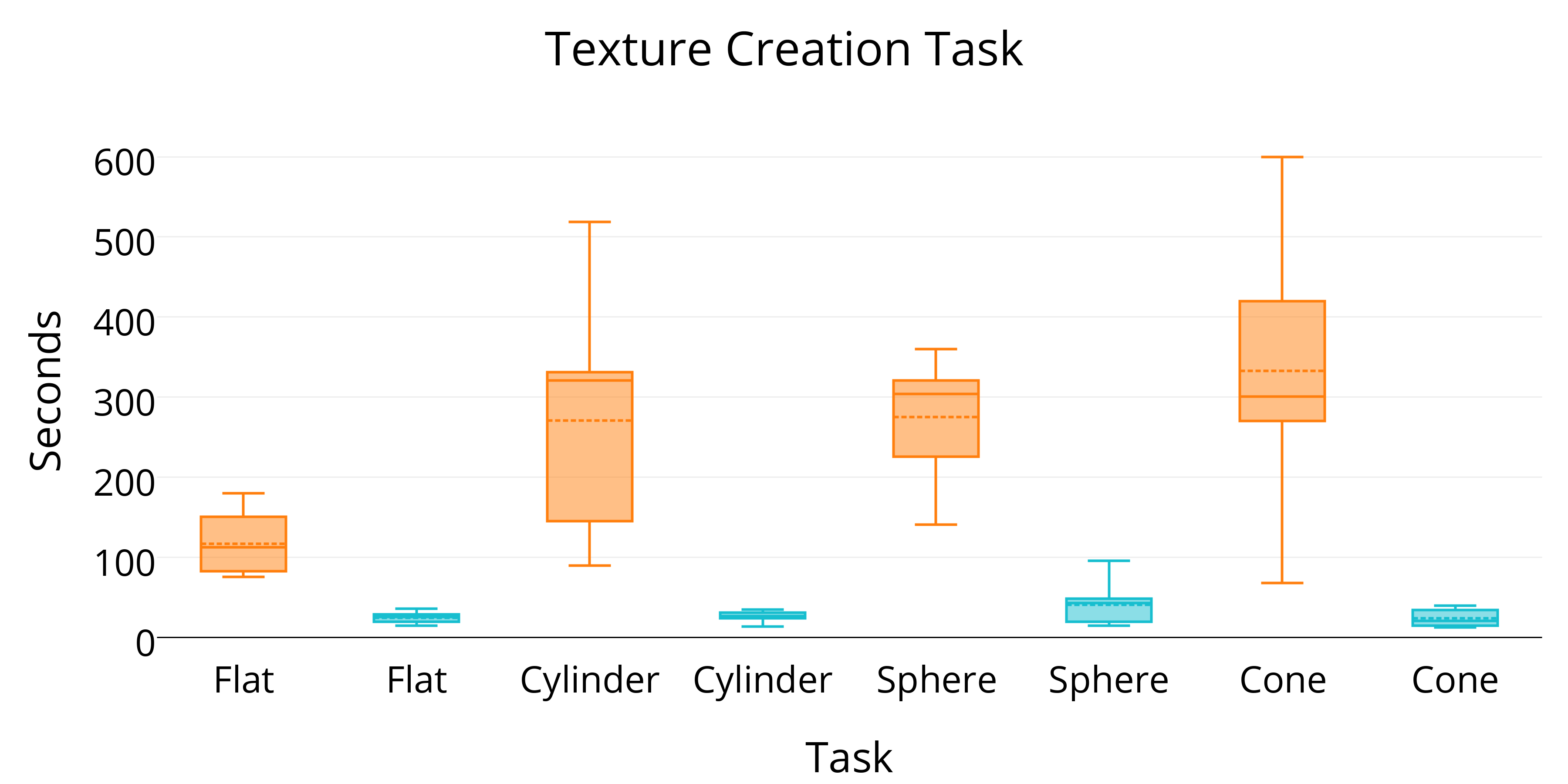}
\caption{Completion time of the texture creation tasks with the reference tool (orange) and Tabby (light blue).}~\label{fig:results}
\end{figure}

\vspace{-1.3cm}
\section{User Evaluation}
To evaluate the usability of Tabby, we conducted a controlled experiment with 7 participants (5: male, 2: female) with some experience in existing CAD tools (Min: 10 months, Max: more than 6 years, Average: 3.2 years).
They self-reported the expertise of 3D modeling tools as 5.3 out of 7 in average.

\textbf{Tasks:} We asked participants to create and modify textures on six different geometries: 1) a flat rectangle; 2) the side of a cylinder; 3) a sphere; 4) the side of a cone; 5) the head of a knight; and 6) the body of the Stanford bunny.
In the creation tasks, the participants were asked to place a small cylinder texture element in a 3x3 grid.
We measured the time by comparing \tool{} with existing CAD tools (we allowed them to choose a tool they felt most comfortable.)

\textbf{Evaluation Results}
With Tabby, participants completed their tasks with 29 seconds on average (Flat: 24 sec, Cylinder: 26 sec, Sphere: 41 sec, Cone: 24 sec).
In contrast, in the reference condition, they needed 248 seconds on average (Flat: 117 sec, Cylinder: 270 sec, Sphere: 275 sec, Cone: 332 sec).
Two participants gave up completing the tasks for sphere and cone surfaces.
Our repeated-measure ANOVA test found significant results for both tools  ($F_{1,6}=48.6$, $p<.01$) and surfaces ($F_{3,18}=6.19$, $p<.01$) as well as their interaction ($F_{3,18}=6.03$, $p<.01$), confirming performance advantages of Tabby.
Participants rated our tool as useful (mean: 6.3 out of 7),  easy to use (5.7), and effective to perform the task (6.4).
The participants liked each feature of Tabby (texture generation: 6.7, auto-completion: 6.1, drag-and-drop manipulation: 5.5, and sketching: 5.7).

\vspace{-0.3cm}
\section{Conclusion}

We present Tabby, an interactive tool to support designing and applying 3D printable textures on an arbitrary complex surface of the existing object.
To minimize repetitive manual efforts, we adopt the \emph{auto-completion} metaphor, which automatically infers the user's demonstration and suggests the possible desired patterns.
To enable this, we develop a series of techniques which infer the user's intention, select the semantic region, and convert 2D shapes into 3D textures.
Our experiment shows that Tabby enables the participants to create and modify 3D texture much faster than conventional tools, especially for complex target surfaces.

\balance

\bibliographystyle{eg-alpha-doi}
\bibliography{references}

\end{document}